\documentclass{amos}
\usepackage[utf8]{inputenc}
\usepackage{amsmath}
\usepackage{amssymb}
\usepackage{graphicx}
\usepackage{biblatex}
\bibliography{bib.bib}

\title{Building Small-Satellites to Live Through the Kessler Effect}

\author{Steven Morad, Himangshu Kalita, Ravi teja Nallapu, and Jekan Thangavelautham \\ Space and Terrestrial Robotic Exploration (SpaceTREx) Laboratory, University of Arizona}
\date{}

\begin{document}

\maketitle

\begin{abstract}
The rapid advancement and miniaturization of spacecraft electronics, sensors, actuators, and power systems have resulted in growing proliferation of small-spacecraft. Coupled with this is the growing number of rocket launches, with left-over debris marking their trail. The space debris problem has also been compounded by test of several satellite killer missiles that have left large remnant debris fields. In this paper, we assume a future in which the Kessler Effect has taken hold and analyze the implications on the design of small-satellites and CubeSats. We use a multiprong approach of surveying the latest technologies, including the ability to sense space debris in orbit, perform obstacle avoidance, have sufficient shielding to take on small impacts and other techniques to mitigate the problem. Detecting and tracking space debris threats on-orbit is expected to be an important approach and we will analyze the latest vision algorithms to perform the detection, followed by quick reaction control systems to perform the avoidance. Alternately there may be scenarios where the debris is too small to track and avoid. In this case, the spacecraft will need passive mitigation measures to survive the impact. Based on these conditions, we develop a strawman design of a small spacecraft to mitigate these challenges. Based upon this study, we identify if there is sufficient present-day COTS technology to mitigate or shield satellites from the problem. We conclude by outlining technology pathways that need to be advanced now to best prepare ourselves for the worst-case eventuality of Kessler Effect taking hold in the upper altitudes of Low Earth Orbit.
\end{abstract}
\section{Introduction}
In 1978, Donald Kessler published a paper titled \textit{Collision Frequency of Artificial Satellites: The Creation of a Debris Belt} \cite{kessler1978collision}. In his paper, Kessler described a future ``debris belt'' caused by cascading collisions between satellites and orbital debris. As a satellite is struck by orbital debris, the satellite would break into many small pieces, becoming debris itself.

Kessler's theory originates from scientists studying the formation of the solar system. Orbital mechanics predicts that orbiting bodies that cross each other's orbits are often unstable and will eventually lead to collision. This collision produces more bodies which cross orbits and will eventually collide. This theory has been used to explain the formation of planetary rings, the asteroid belt, and even formation of the planets \cite{kessler2010kessler}. Kessler used it to describe Earth's artificial satellites.

\begin{figure}
    \centering
    \includegraphics[width=0.5\linewidth]{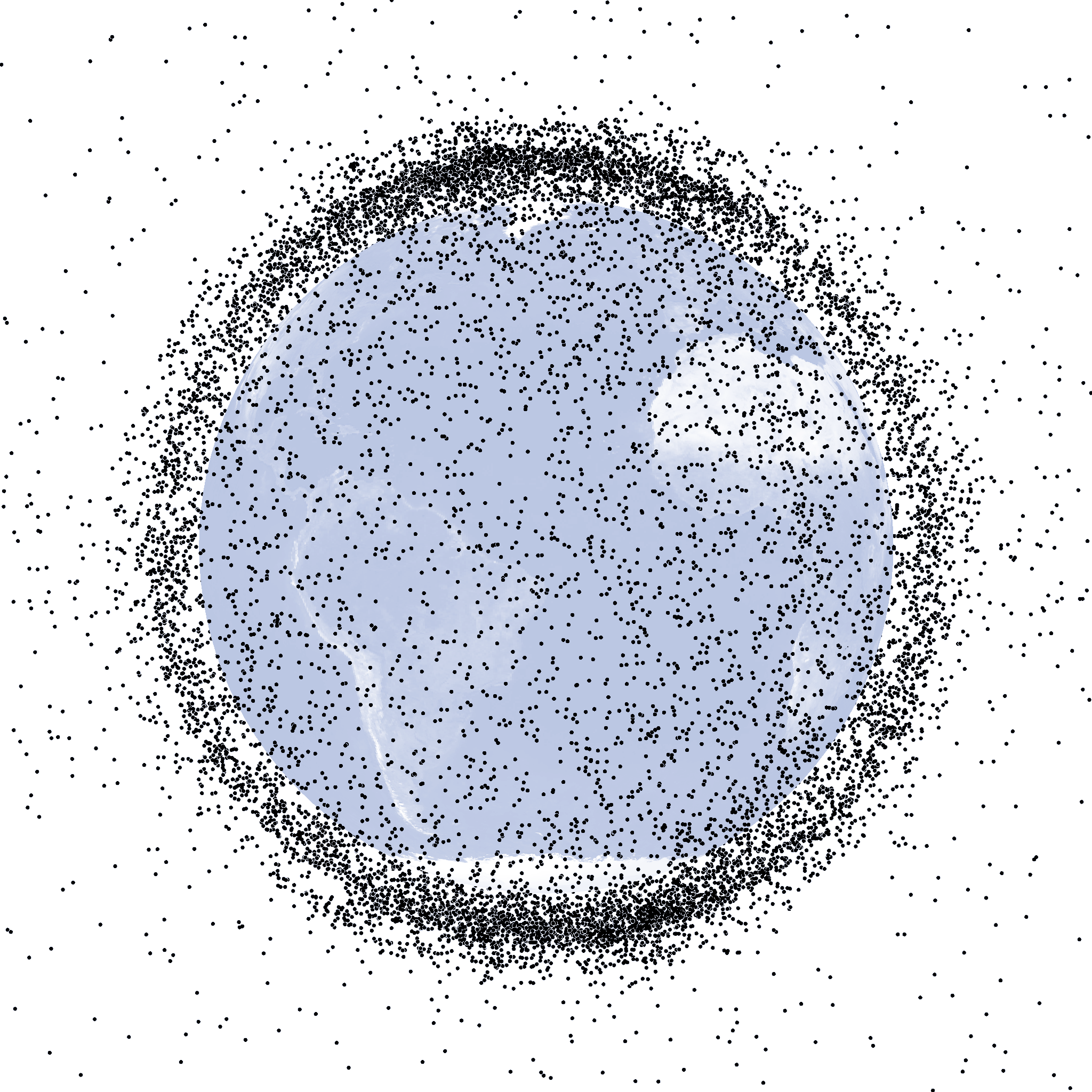}
    \caption{A picture of debris in low Earth orbit from 2009. Since then, the problem has only gotten worse. \cite{2009}}
    \label{fig:debris}
\end{figure}
The Iridium-Kosmos satellite collision in 2009 and resulting debris cloud has given us a small taste of the Kessler effect. The anti-satellite test by the Chinese government in 2007 generated over 35,000 pieces of debris \cite{kelso2007analysis}, further accelerating the pace of the Kessler effect. Now, with plans for multiple megaconstellations in LEO we must begin thinking about how to build satellites to survive a cascading collision scenario.

\subsection{The Effect of Megaconstellations}

The term ``megaconstellation'' is a relatively new word. It describes a synchronized orbital arrangement of several hundred satellites. Over the past decade there has been a drastic decrease in launch costs and spacecraft hardware, partially due to the rise of CubeSats, small satellites built using mostly Commercial Off The Shelf (COTS) parts. It is cheaper than ever to build and launch satellites, and this is changing how commercial entities think about space. SpaceX, Amazon, OneWeb, and TeleSat are all planning to launch megaconstellations in the next few years.

SpaceX plans to launch the first of almost 12,000 satellites this year \cite{foust2018spacex} as part of their Starlink network. This megaconstellation serves to provide satellite internet throughout the world. These will be launched into three orbital altitudes, 7,500 satellites at 340 km, 1,600 satellites at 550 km and 2,800 satellites at 1,150 km.

OneWeb will be launching a 650-satellite megaconstellation with a similar purpose to StarLink -- to provide satellite based internet access to the world. The OneWeb constellation altitude is set to be 1,200km \cite{foreman2017large}.

TeleSat, like others, plans to launch a megaconstellation for communications purposes. They plan to launch their constellation at between 1,000 and 1,200 km altitude \cite{del2019technical}.

The Amazon Kuiper megaconstellation will consist of 3,236 satellites to provide broadband internet to the world. The Kuiper satellites will reside at altitudes between 590 and 630 km.

Radtke et al. published a study on the effects of the OneWeb constellation. They found that at 800 km altitude, each satellite has a 69.35\% chance of colliding with a 3 cm object or larger over their short orbital lifetime, using the current debris flux at that altitude \cite{radtke2017interactions}. They found that a \textbf{single collision} increases the debris flux for other satellites in the constellation by a \textbf{factor of nine}. Just one megaconstellation could lead to cascading collisions, resulting in large changes in orbital energy of the debris. These changes in energy result in changes in orbital parameters, spilling debris over into other orbital regimes/altitudes. With many different megaconstellations launching over the next decade, the chance of cascading collisions leading to environmental catastrophe greatly increases.  

In the case of a catastrophe where space debris dominates LEO, we must be ready for this new eventuality. Satellites now serve an integral role in daily life, and future designs need to survive (mitigate) such events. In this paper, we look at the methods available today to help a small satellite survive the realization of the Kessler effect.

\subsection{Satellite Survival Methods}

We break the satellite survival methods down into three categories:
\begin{itemize}
    \item Remote Avoidance
    \item Local Avoidance
    \item Passive Shielding
\end{itemize}

Avoidance represents using onboard actuators to execute maneuvers to avoid debris. In this case, the debris is seen and tracked. Action must be taken on the part of the spacecraft to avoid colliding with a piece of debris in the future.

The term ``remote'' in remote avoidance denotes the source of conjunction data. In the remote case, data likely comes from ground based assets such as radar or telescopes. Additionally, avoidance maneuvers may be produced on the ground or produced onboard and referred for execution from the ground. Remote avoidance works for large, ground-trackable pieces of debris that are far away from the satellite. The joint DoD-NASA effort to catalog orbital debris cannot track objects smaller than 5cm \cite{size}.


Local avoidance utilizes an on-board data source to generate maneuvers for collision avoidance. Instead of ground-based radar, on-board sensors are used. In this case, the satellite has little time to wait for ground input, so all avoidance happens onboard the satellite. The benefit of this type of avoidance is it does not rely on the ground to dodge debris. Similar to remote avoidance, local avoidance works for big pieces of debris at large distances. However, it also works for small pieces of debris and small distances. Although small, these pieces of debris can still be mission-ending.

Finally, for particles too small for active avoidance, passive satellite shielding can be used. An example of a passive shield is a Whipple shield \cite{christiansen1993design}. Whipple shields have been demonstrated for shielding on numerous satellites and even human-class spacecraft, such as the International Space Station \cite{christiansen2009space}. For common orbital energies, Whipple shields can only protect against debris one centimeter or smaller \cite{krag20171} \cite{NAP5532}.

\subsection{Motivation}
There is ample literature on remote avoidance and passive shielding. In fact, these systems have been proven on many objects currently in orbit. Less has been done on local avoidance, where we decide to focus this paper. Local avoidance sits in the space between passive shields and remote avoidance. Local avoidance exists to mitigate debris too small to be tracked by ground systems, yet large enough to penetrate Whipple shields. 


\section{Method}
The debris avoidance problem consists of two parts. The first part is detection, we must use onboard sensors to detect debris along the satellite trajectory. The second part is maneuvering to avoid the detected debris.
\subsection{Detection Methods}
We can use either passive or active sensors for detection. Active sensors, like radar, consist of both transmitter and receiver. The transmission effectiveness scales with the amount of power available. For radar, an increase in power corresponds to an increase in range. Small satellites are power constrained, which makes active sensors less effective and less desirable.

\begin{figure}
    \centering
    \includegraphics[width=0.5\linewidth]{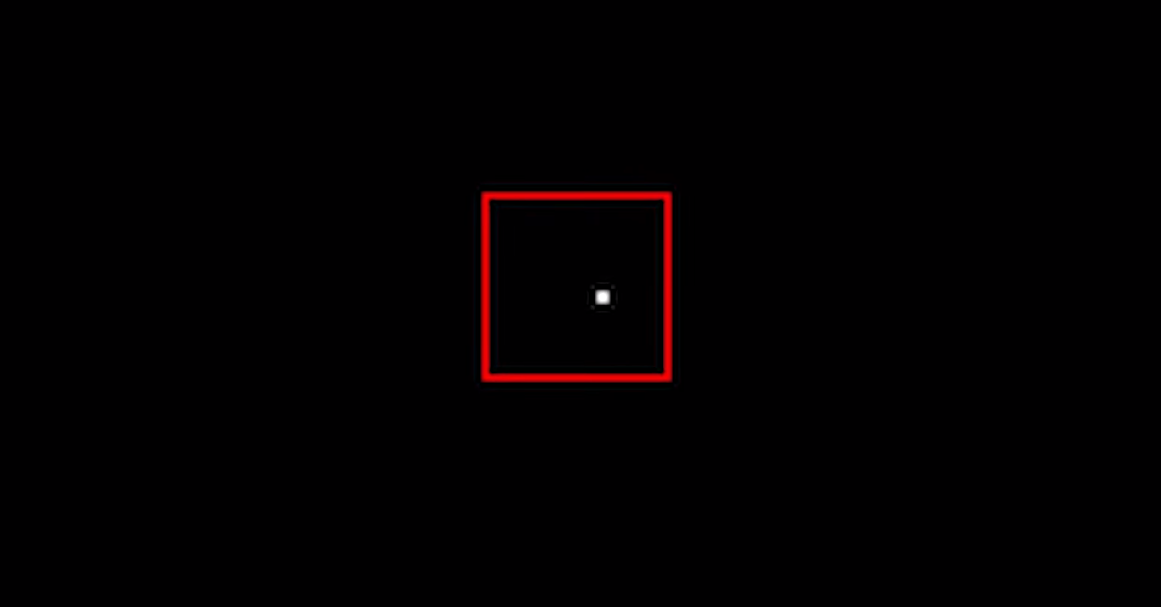}
    \caption{Picture from a RIM-161 SM3 thermal infrared camera, tracking a target in space \cite{missile}. The difference between hot and cold pixels is sufficient for detection, no advanced computer vision algorithms are required.}
    \label{fig:tracking}
\end{figure}

Passive sensors, such as cameras, do not radiate power like active sensors. They consist of only a receiver, and rely on energy emitted from the environment. Optical cameras can only reliably detect objects during the day, at night debris does not emit energy in the optical wavelengths. However, debris constantly emits blackbody radiation in thermal infrared wavelengths. Anti-satellite missiles like the Raytheon RIM-161 use thermal infrared tracking for terminal stage guidance (Fig. \ref{fig:tracking}) \cite{kopp2008theatre}, because orbital objects are constantly emitting blackbody thermal infrared radiation. These anti-satellite missiles are also volume and power constrained just like small satellites, the difference being anti-satellite missiles are trying to hit a target travelling at very high speeds while small satellites are trying to miss a target. The drawback to thermal infrared detection is limited range. Anti-satellite missiles are radar guided for a large duration of their flight. Only once they get within tens of kilometers of their target does the kinetic warhead separate and utilize thermal infrared guidance\cite{kopp2008theatre}. This means our avoidance system will need to react extremely fast.

\begin{figure}
    \centering
    \includegraphics[width=0.5\linewidth]{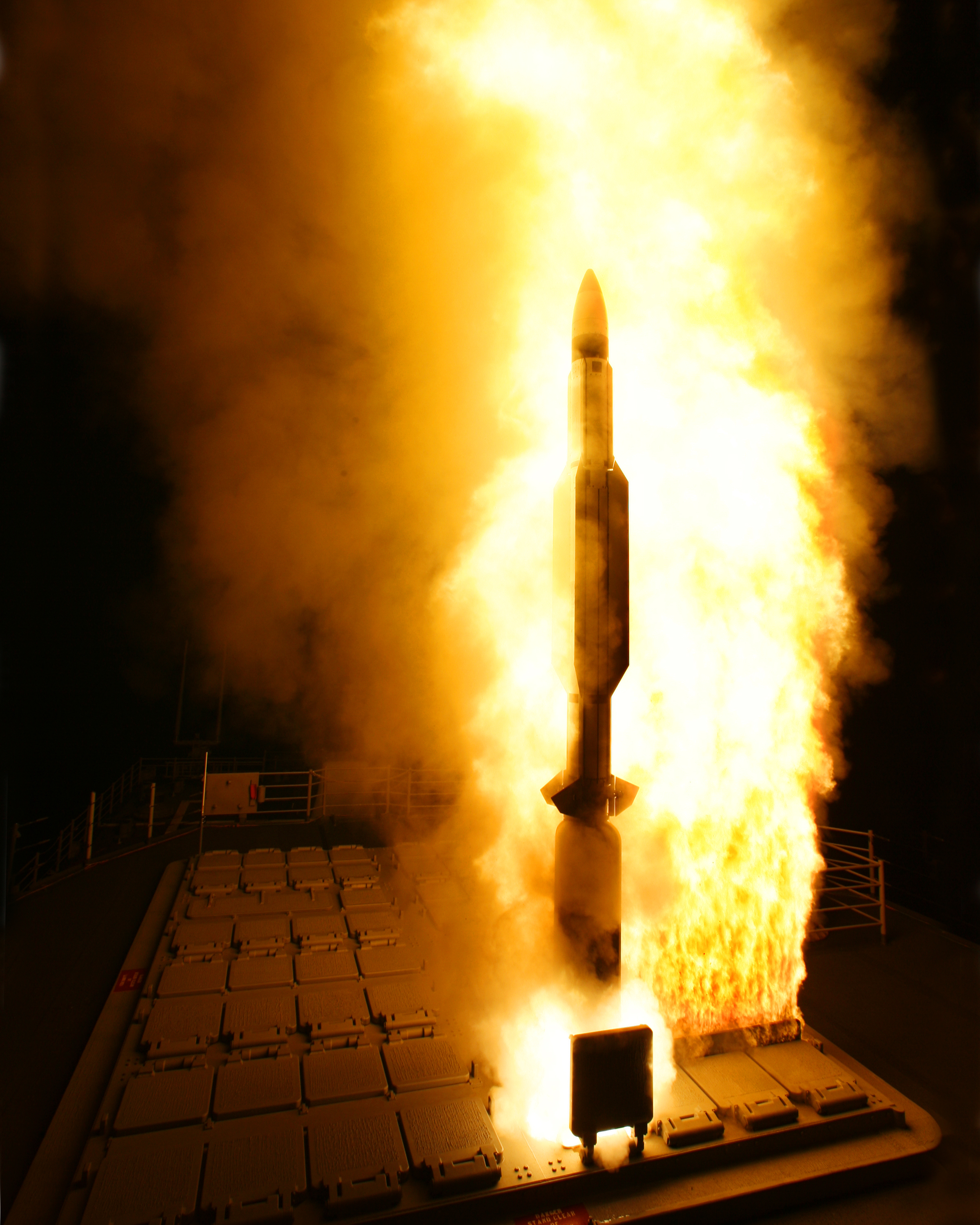}
    \caption{Launch of a RIM-161 SM3 anti-satellite missile \cite{schulze_2007}. We utilize systems similar to the RIM-161 SM3 kinetic warhead for debris detection and avoidance maneuvers.}
    \label{fig:ircam}
\end{figure}

\subsubsection{Probabilistic Detection Algorithm}

We can model the detection probabilisticly. At large distances, the object should only take up one pixel. We model each pixel as ``hot'' $p(H)$ or ``cold'' $p(\neg H)$ as the product of signal $s$ and some noise $\eta$ 

\begin{align}
P(H) = P(H|s,\eta)
\end{align}
We model this over $n$ frames
\begin{align}
P_{n}(H) &= \prod_{i=0}^n  P\left(H|s_i,\eta_i\right)
\end{align}
Assuming the noise per pixel is uniformly distributed about the true pixel value $\mu$

\begin{align}
P(H|s,\eta) \sim \mathcal{N}(\mu, \sigma)
\end{align}
In the limit, we can expect the noise contribution to go to zero and the probability approach the true value. This is because the product of Gaussian distributions becomes the Dirac delta function in the limit
\begin{align}
&\lim_{n \to \infty} P_n(H|s,\eta) = \delta(X-\mu)\\
\end{align}
We relax the single-pixel assumption to a small neighborhood of pixels, which allows the object to move across pixels, but still remain tracked.

Now, we can track $P(H)$ for a neighborhood around each pixel over $n$ frames to separate debris from noise. If we have a hot pixel that remains in the previous hot pixel's neighborhood for $n$ frames, it is likely we have detected incoming space debris. One of the benefits of using a thermal infrared detector is that we only have to worry about hot and cold pixels. We do not have to do any object recognition (Fig. \ref{fig:ircam}). With the known spacecraft attitude, we can compute the position of the debris in the camera plane, and extrapolate the trajectory of the debris. At these small orbital scales we can assume the trajectory of both the object and debris are linear. If the debris lies along the velocity vector of the satellite, the satellite must take action to prevent collision.

\subsection{Avoidance Methods}
When looking for trajectory-altering maneuvers, we need a type of propulsion system that is high-thrust, quick to activate, reliable, and lightweight. The high-thrust requirement immediately rules out many forms of electric propulsion. Petroleum-based fuels like RP-1 and oxygen provide large amounts of thrust, but require seconds to ignite, which rules them out for extremely fast reactions. Hypergolic fuels and cold gas thrusters fit the bill, but have relatively low I$_{sp}$ when compared to solid-fuel rockets. Solid-fuel motors check all the boxes, they activate in a fraction of a second, are reliable, high-thrust, and lightweight. The issue with conventional solid motors is they are one-time use. By carrying multiple conventional solid motors, our avoidance system can activate many times.  Alternately, there are solid gel rockets that can be electrically activated or stopped.

\begin{figure}
    \centering
    \includegraphics[width=0.5\linewidth]{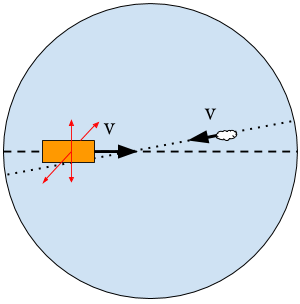}
    \caption{The orange satellite and white debris are on a collision course with a shallow difference in inclination. Since we do not have time to reorient the spacecraft, thrusting in the plane denoted by red vectors provides the best chance of survival.}
    \label{fig:diag}
\end{figure}

\begin{figure}
    \centering
    \includegraphics[width=0.5\linewidth]{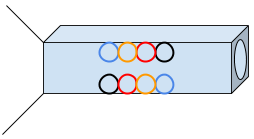}
    \caption{A 3U CubeSat with color-coded two-thruster groups that fire in unison to move the spacecraft away from debris. The thruster groups are symmetric about the center of mass, to reduce the net moment created by the thrusters.}
    \label{fig:placement}
\end{figure}

We place multiple conventional solid motors along faces orthogonal to the direction of travel. In other words, the thrusters move the satellite in the plane normal to the satellite velocity vector (Fig. \ref{fig:diag}). Since the majority of impacts happen in the direction of travel \cite{vance2013value}, orthogonal displacement provides the best chance of collision avoidance. Groups of two thrusters are placed opposite the center of mass of the spacecraft to minimize the moment generated by the thrusters (Fig. \ref{fig:placement}).

\section{Analysis}
In this section, we systematically analyze the performance of the systems discussed in the previous section.

\subsection{Detection System}
\begin{figure}
    \centering
    \includegraphics[width=\linewidth]{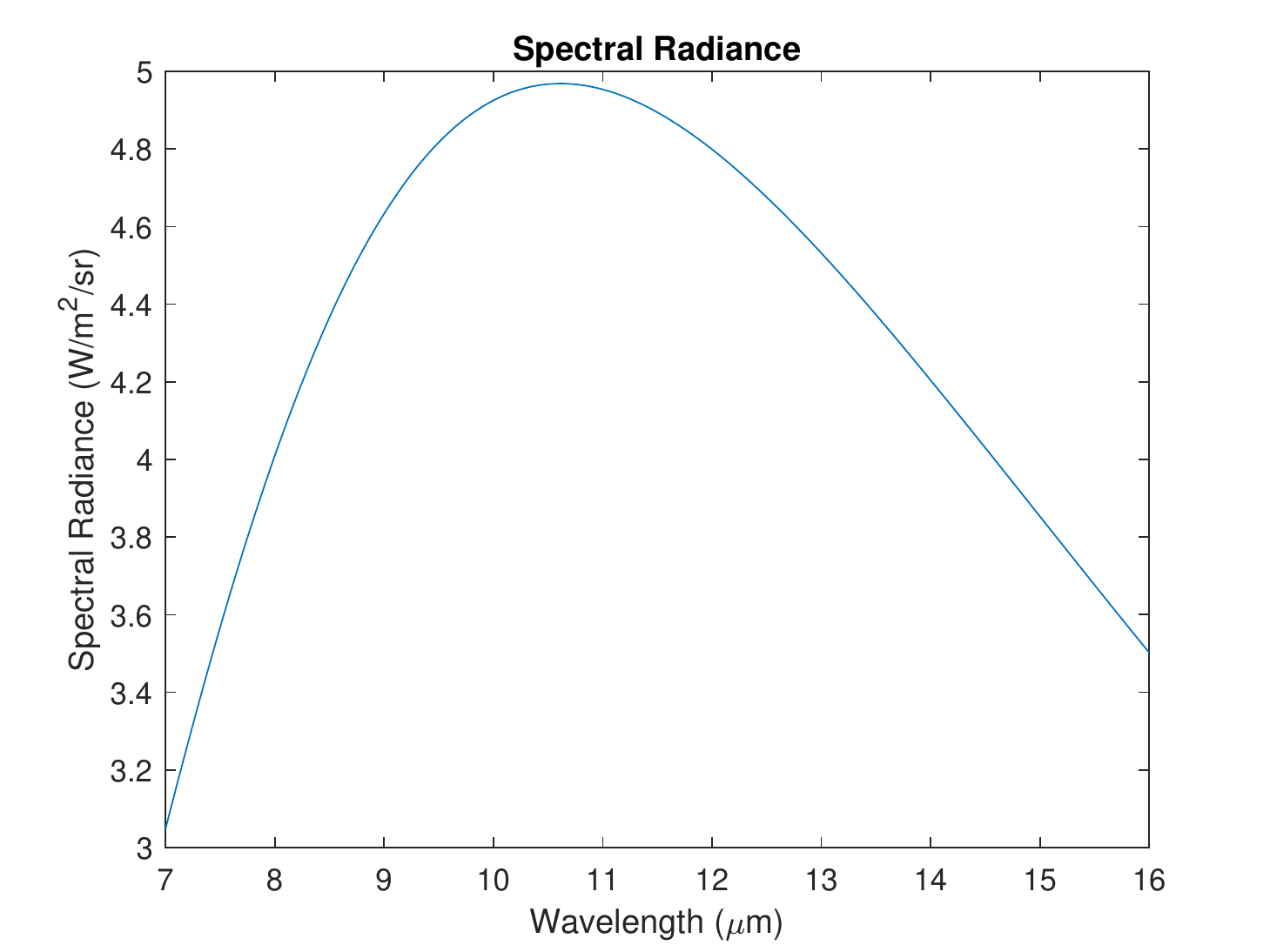}
    \caption{Spectral radiance curve $\epsilon P(\lambda)$ for an object emitting blackbody radiation at 273K. We use an emissivity of $\epsilon = 0.8$. $\epsilon P(\lambda)$ can be integrated to find the total detectable radiation emitted by a piece of space debris.}
    \label{fig:radiance}
\end{figure}

According to the Monte-Carlo simulation in \cite{vance2013value}, we expect that the majority of collisions happen head-on or retrograde with a shallow relative inclination to a spacecraft. A single forward-facing infrared detector with a sufficient field of view should be able to detect a majority of collisions. We use a theoretical thermal infrared sensor that is sensitive to wavelengths from seven to sixteen microns. We assume debris temperature of 273K and generate a spectral radiance curve $P(\lambda)$ that denotes emitted blackbody radiation (Fig. \ref{fig:radiance}). 

Integrating this curve, we find the energy emitted per surface area in the detector-sensitive wavelengths. Typically, debris one centimeter and larger has been determined dangerous \cite{NAP5532} \cite{vance2013value}, so our theoretical piece of debris is spherical with a radius of half a centimeter

\begin{equation}
    r = 0.5\text{cm}
\end{equation}

which provides a surface area of

\begin{equation}
    s = 4 \pi r
\end{equation}

We use an emissivity of 0.8 for the piece of debris. 
\begin{equation}
    \epsilon = 0.8
\end{equation}

For reference, the emissivity of anodized aluminum is 0.78 and the emissivity of clear acrylic plastic is 0.94. This gives us a radiant intensity of 

\begin{equation}
    I = \epsilon s \int{P} d\lambda
\end{equation}

\begin{figure}
    \centering
    \includegraphics[width=\linewidth]{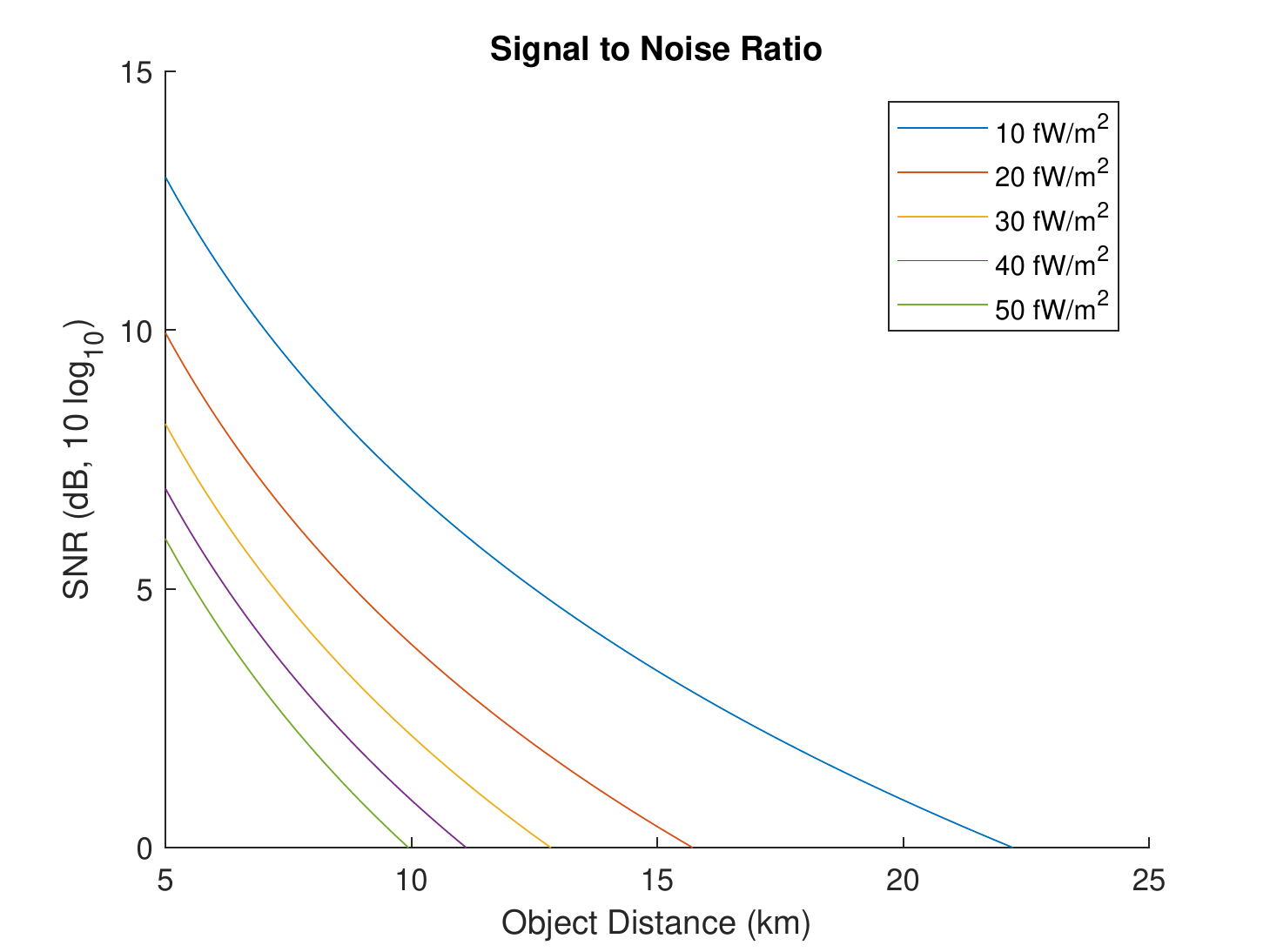}
    \caption{Signal to Noise Ratio in decibels for various NEIs. The short detection distance means fast reactions are required to avoid collisions. Various processing algorithms can be used to glean signal from lower SNRs, but once the decibels reach zero, signal fades into the background noise. Even with 50 femtowatts per square meter, we can theoretically detect debris at 10 km away. For comparison, the MSX spacecraft IR camera band centered at $\lambda= 8.28 \mu$m produced an NEI of 7 fW/m$^2$ \cite{price2004spectral}.}
    \label{fig:snr}
\end{figure}

We compute the intensity per surface area at the camera aperture to find the irradiance cast by the object on the camera sensor. As expected, the signal is weak but as we will show it is still detectable at kilometer scales. Figure \ref{fig:snr} shows the expected Signal to Noise Ratio (SNR) as a function of debris distance from the satellite. Sensor noise is converted into an irradiance value known as Noise Equivalent Irradiance (NEI). NEI allows for an apples to apples comparison of incoming signal and sensor noise.

\begin{figure}
    \centering
    \includegraphics[width=\linewidth]{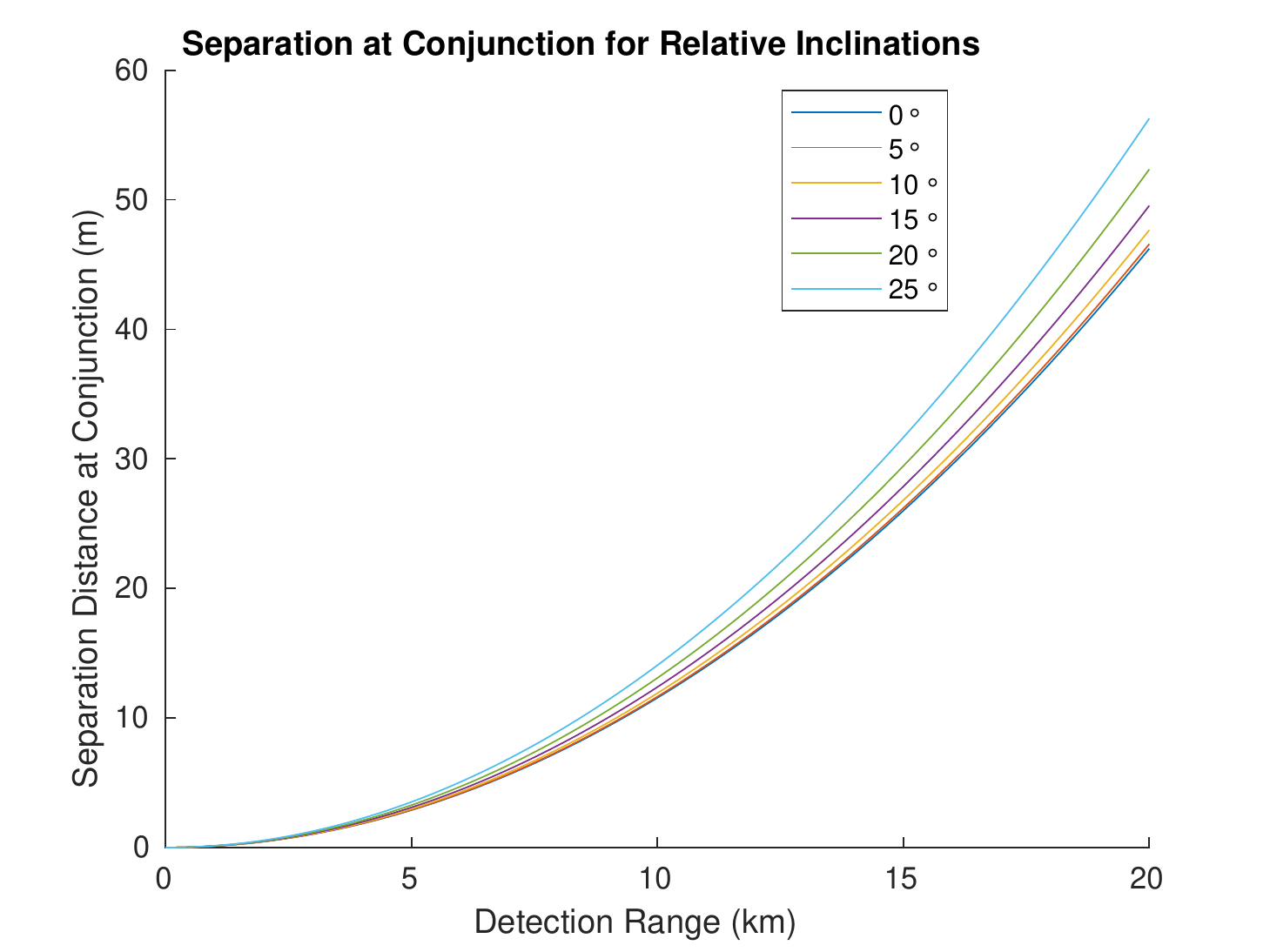}
    \caption{Separation for various relative inclination orbits if the thruster is fired at debris detection. A CubeSat travelling in a circular prograde orbit has a relative inclination to debris travelling in a retrograde circular orbit. The orbital altitude is 1000 km. Zero degrees corresponds to debris in a retrograde orbit in the same plane as the CubeSat. Larger relative inclinations result in slighty more time to maneuver and reduce the probability of a collision.}
    \label{fig:sep}
\end{figure}

It is unclear exactly how much NEI a CubeSat form factor IR sensor would produce. The Midcourse Space Experiment (MSX) launched in 1996 had a thermal IR camera with an NEI of $10^{-13}$ to $10^{-15}$ Wm$^{-2}$ depending on the band \cite{price2004spectral}. Recent efforts to produce CubeSat-sized thermal imagers have seen progress, and sensors have been developed and built \cite{thompson2019mmt}. Recent papers have shown that low-noise thermal infrared sensors similar in quality to the THEMIS instrument on the Mars Reconnaissance Orbiter (MRO) can fit inside a 3U CubeSat today \cite{puschell2014uncooled}. Reliable, space-grade, low-noise infrared sensors are a critical pathway for local debris detection.



\subsection{Maneuver System}
Visual detection of debris is just half the challenge. Once the debris is detected we must maneuver to avoid it. The maneuver system uses solid rocket motors to move the spacecraft to safety. The Aerotech G339N-P solid motor provides roughly 110N-s of impulse over 0.4s. By reducing the length of the thruster, we can increase the number of carried motors and ``manuever charges''. Scaling the length of the actual thruster to one quarter produces a theoretical thruster that provides 27.5N-s over 0.1 seconds. Mounting two motors symmetrically over the center of mass as in figure \ref{fig:placement} results in 55N-s of impulse over 0.1 seconds. We simulate the motor performance on a 4 kg, 3U CubeSat platform.

We calculate the relative orbital speed of the satellite and debris using circular orbits, with varying relative inclinations. We simulate the motor impulse over 0.1s after debris detection to produce figure \ref{fig:sep}. Note that this is one of the worst-case scenarios -- retrograde debris with an identical inclination provides very little time for avoidance. Nonetheless, we can create a sizable separation distance from the debris in very little time, and that changes in relative inclination make little differences at this range.


\subsection{Conceptual Cubesat}
\begin{figure}
    \centering
    \includegraphics[width=4.0in]{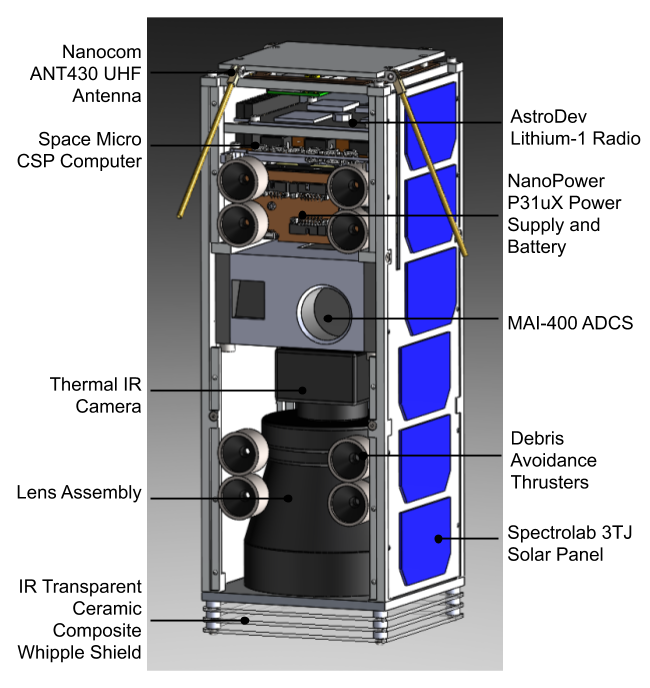}
    \caption{A conceptual CubeSat design that demonstrates our space debris avoidance system. The system is based off of the SWIMSat CubeSat we previously designed for meteor monitoring \cite{morad2018orbit,hernandez2016swimsat}. The IR transparent ceramic Whipple shield serves to protect the front of the spacecraft from small debris, while the IR camera and thruster system allows fast avoidance of debris.}
    \label{fig:cubesat}
\end{figure}

Using a theoretical CubeSat thermal infrared camera and the maneuver system, we present a CubeSat concept to demonstrate the packaging of the system (Fig. \ref{fig:cubesat}). One issue is the camera must be front facing, but that is also where the majority of small impacts occur. These small impacts could destroy the camera or lens assembly. Kevlar and Nextel Whipple shields have a disadvantage in that they would obscure the camera.

Fortunately, missile research and development again saves the day. Heat-seeking missiles have a strong IR-transparent dome to protect the delicate IR sensor and related electronics. Over the past few decades, IR transparent ceramics have progressed significantly for this purpose \cite{stefanik2007nano}. The toughness of these ceramic composites also makes them great Whipple shields \cite{silvestrov1999investigation}.

\section{Conclusion}
We discussed the Kessler effect, and how recent trends such as megaconstellations may exacerbate this effect. As launch costs decrease and space becomes commercialized, cascading satellite collisions are bound to become increasingly more likely. We need to start designing satellites that can survive in orbits densely populated with debris and other forms of space traffic. Whipple shields for passive shielding and ground-based conjunction detection and avoidance is already a well-studied problem. However, these systems cannot mitigate debris too small for detection, but too big for shielding. We proposed a method named local avoidance -- where a small spacecraft can detect an upcoming collision and avoid it without any input from the ground. Power constraints leave no room for high powered active sensors such as radar, so short range passive thermal infrared sensors like those found in anti-satellite missiles are used instead. These sensor systems can detect debris tens of kilometers away. If the debris is traveling retrograde, that gives us seconds to react. Immediately after detection, laterally mounted high-thrust solid-fuel rocket motors provide a fast and rough push out of harms way. This detection and avoidance system could prove useful to future small satellites trying to avoid debris fields. 

\printbibliography
\end{document}